\documentstyle[11pt,aaspp4,flushrt,epsf]{article}
\renewcommand{\reference}{\bibitem{dummy}}
\renewenvironment{references}{\begin{thebibliography}{}}{\end{thebibliography}}

\begin{document}

\title{Bringing light to galaxy formation simulations\altaffilmark{1}}
\altaffiltext{1}{Slightly expanded version of article to appear in {\it Dark
and visible matter in galaxies}, editors M.\ Persic and P.\ Salucci, ASP
Conference Series.}

\author{Sergio Gelato}
\affil{SISSA, 2--4 via Beirut, I--34013 Trieste, Italy}
\author{Fabio Governato}
\affil{Department of Astronomy, FM--20, University of Washington,
Seattle WA 98195, USA}

\begin{abstract}
Models of galaxy formation ultimately aim at reproducing the {\em
observed} properties of galaxies.
We report on work in progress to predict luminosities, colours and
morphologies of field objects of various masses through
$N$-body~$+$~SPH simulations.
We describe our method, illustrate the effects of varying the star
formation assumptions, show some preliminary results, draw
encouragement from their strengths as well as lessons about what
improvements are most needed.
\end{abstract}


\section{Introduction}
\label{s:intro}

$N$-body simulations are a well established tool to test
theories of structure formation in cosmology.
Within their domain of validity---whenever gravity is the single
dominant physical process, as may be the case on sufficiently large
scales for a universe dominated by cold dark matter (CDM)---they yield
robust predictions about the numbers, size distribution and clustering
properties of mass condensations in any given cosmological model.
Their potential for new useful discoveries is not yet exhausted, as
the recent work of Navarro et al (1996, hereafter NFW) illustrates.

Unfortunately their predictions about the {\em mass} distribution
don't easily lend themselves to a comparison with observations,
which only detect {\em radiation} sources.
It is now very clear that on scales as small as individual galaxies
the light is a poor tracer of the mass.
Connecting the two requires further modelling, whether one adopts the
``inverse'' approach of deducing
the mass distribution from the properties of the
luminous material or the more ``direct'' one of adding to the $N$-body 
simulations the main physical (thermodynamical,
chemical, nuclear) processes that ``give light'' to galaxies.
In both cases we are faced with huge uncertainties stemming from our
lack of understanding of the key processes of star formation and
feedback to the interstellar medium.
The inverse approach has the additional drawback of discarding some
observational information; this becomes increasingly wasteful as the
quality and quantity of observations improves.
At the same time, advances in computing power and in numerical
techniques are making direct simulation ever more practical.
The pioneering work of Katz (1992), 
Navarro \& White (1993, hereafter NW), 
Steinmetz \& M\"uller (1995), 
to cite but a few, serves as a springboard for
further exploration of this area of crucial interest for cosmology.

To predict galaxy luminosities and spectra, we must keep track at the
very least of the cooling and condensation of primordial gas, of the
formation and evolution of stars, of the stars' feedback on the state
and composition of the surrounding gas.
Our ignorance of the details of these processes, and Occam's razor,
call for very schematic models with few free parameters, based
mainly on simple conservation laws.
We assume that star formation occurs predominantly in cold, high
density, collapsing gas and that the energy of supernova winds
eventually heats the surrounding gas.
A test bed for these ``recipes'' is provided by the recent wave of
semi-analytic models of galaxy formation (reviewed by Frenk at this
conference).
We can also draw useful lessons from a line of research (exemplified
by Pardi et al 1995) that relies on very detailed chemical evolution models
to deconvolve the formation history of observed stellar populations in
the Milky Way.
While valuable as sources of observational constraints and analysis
techniques, such models don't readily help us choose among
cosmological scenarios.
And like the semi-analytic models, they treat only schematically
the spatial
dependence of the star formation and feedback processes, the explicit 
inclusion of which is one of the $N$-body method's greatest strengths.

This method is still in its infancy, which means we must
learn more about the validity of our model assumptions before
we can draw firm conclusions about which cosmogony is the correct one
for our universe.
But in the long run we hope and expect direct simulations to
shed light over many outstanding questions, such as:
the origin of the morphology-density relation;
the overabundance of dwarf galaxy halos in CDM;
the relation between total galaxy mass and the shape of the rotation curve;
the nature of the faint blue galaxies;
the ``nature vs. nurture'' hypothesis for the origin of elliptical
galaxies;
the origin of low surface brightness galaxies.

\section{The method}
\label{s:method}

\subsection{Initial conditions}
\label{s:init}

We start with a Gaussian random field with an $\Omega=1$ cold dark
matter (CDM) power spectrum normalized to $\sigma_8=0.7$ at the
present epoch on a periodic three-dimensional grid of 256 nodes and
32~Mpc on a side. 
We adopt $H_0=50\,{\rm km}\,{\rm s}^{-1}\,{\rm Mpc}^{-1}$, so that
$\sigma_8^2$ is the mass variance within a sphere of radius 16~Mpc.
This choice of parameters may not yield the best fit to observations,
but it is a standard benchmark for this class of simulations.

The density field is first evolved using a particle-mesh code, and
interesting halos, defined as connected high density
($\delta\rho/\rho>57$) regions, are selected from the final state.
In this report we present results for two halos, code-named B6 and B9, with
respective circular velocities $V_c$ of 260 and 120~km~s$^{-1}$.
These halos were chosen to lie outside clusters, and are thus
expected to correspond to spiral rather than elliptical galaxies.
The initial location of each halo is then resampled with equal numbers of
gas and dark matter particles (3000 to 4000 each) in a sphere large
enough to hold the protohalo.
Surrounding concentric spheres are sampled with
collisionless particles only, at progressively coarser resolution,
to produce a realistic tidal field.
The particle positions and growing-mode velocities are obtained by
perturbing a $128^3$ (B6) or $256^3$ (B9) lattice according to the
Gaussian random field realization mentioned earlier, which already
includes all the short-wavelength power we need.

\subsection{SPH simulations}
\label{s:SPH}

Smoothed Particle Hydrodynamics (SPH) is reviewed in Monaghan (1992).
Our implementation is an evolution of that of NW.
The most significant additions are:
a heat source that mimics a photoionizing UV background (Navarro \&
Steinmetz 1996);
a few alternative star formation recipes;
bookkeeping of the mass of heavy elements produced in stars and
returned to the gas.

The gravitational softening, 1.25~kpc (B9) to 2~kpc (B6) for the gas
and twice as much for the DM, allows us to start the simulations at
$z=39$ and $z=29$ respectively. Our minimum time step, dictated by
accuracy and stability requirements, is about $2\times 10^5$~yr, making
a complete simulation rather time consuming.

The cooling function incorporates Bremsstrahlung and atomic line
cooling for a primordial Mixture of H and~He, but no H$_2$ cooling.
For the time being we also neglect the enhancement to the
Bremsstrahlung as heavy elements are added to the gas.

In some runs we include a heating term from an ultraviolet photon
background, such as might be produced by quasars and
Population~III stars. This background modifies the ionization
equilibrium of the gas, particularly at low densities, reducing the
cooling rate as well as providing additional energy.
We follow existing practice by choosing a flux
\begin{equation}
\label{q:Jnu}
J_\nu(z) = J_{-21}(z) \times 10^{-21}
\left(\nu_L\over\nu\right)^\alpha
\,{\rm erg}\,{\rm s}^{-1}\,{\rm cm}^{-2}\,{\rm sr}^{-1}\,{\rm Hz}^{-1}
\end{equation}
with $\alpha=1$ (more appropriate%
\footnote{Perhaps even a little too hard, given a mean quasar
spectral index $\alpha\sim1.5$ and the presence of intervening
Ly~$\alpha$ absorbers. We merely followed the example of Katz et al
(1996) and Navarro \& Steinmetz (1996). Given the crudeness of other
aspects of our method, this discrepancy should be no particular cause
for concern.}
 to a non-stellar origin of the photons) and
\begin{equation}
\label{q:J21}
J_{-21}(z) = {1-e^{z-8} \over 1 + \left[5/(1+z)\right]^4 }
\end{equation}
for $0\le z \le 8$.
Our normalization agrees well with observational determinations of the UV
flux (Giallongo et al 1996).

We allow stars to form in regions where the gas flow is convergent and
the gas density exceeds $7\times
10^{-26}\,{\rm g}\,{\rm cm}^{-3}$. A similar value of
$1.67\times10^{-25}\,{\rm g}\,{\rm cm}^{-3}$ has been advocated by
Kennicut (1989) on observational grounds.
This threshold guarantees that the gas can cool to $10^4$~K in much
less than the local dynamical time $t_d=(16G\rho/3\pi)^{-1/2}$,
contracting isothermally afterwards.
Where these prerequisites are satisfied, the gas is
turned into stars at a rate $d\rho_{\rm gas}/dt=-\rho_{\rm gas}/t_d$.
We discretize the process by waiting a time $t_d|\ln(1-\varepsilon)|$
before turning a fraction~$\varepsilon$ of an eligible gas particle
into a new particle of coeval stars; 
our conditions for star formation
must remain satisfied throughout this finite time interval.
The value of~$\varepsilon$ sets our time resolution of the star
formation event. 
In this particular set of simulations we found adequate convergence
for~$\varepsilon\la 0.15$.

For comparison, we tried raising the density threshold, either
explicitly to the Kennicut value or by requiring that the gas be Jeans
unstable on the smallest resolved scale: $t_s \equiv h/c_s > t_d$,
where $c_s\sim 12\,{\rm km}\,{\rm s}^{-1}$ is the sound speed and $h$
the local gas resolution (taken as the greater of the SPH
smoothing and gravitational softening lengths).
This amounts to a resolution-dependent density threshold $\rho >
m^{-2} (3\pi c_s^2/16G)^3$, where $m$ is the mass of a gas resolution
element.
Such behaviour can be useful to counteract the
effect that lower resolution implies averaging of the density over
larger regions.
In our B9 runs the implied threshold is higher than Kennicut's.

Once formed, a star particle begins to lose mass to the surrounding
gas particles at every integration step according to standard stellar
evolution models.
(For this and other purposes we assume a Salpeter IMF between 0.1 and
40~$M_\odot$.)
Thermal energy is dumped into the gas as part of the same process.
Pending further tests, we do not add bulk kinetic energy to the gas.

We keep track of the mass in elements heavier than Helium
(``metals'').
This is initially zero for all gas particles, and stars inherit the
metal fraction of their parent gas particles.
The material they feed back to the gas contains this same fraction,
enriched by the mass newly produced in supernova progenitors
according to the following fit to Woosley and Weaver's (1995) data:
\begin{equation}
M_Z(M) =
2.935 - 0.65 M
+ 4.59 \times 10^{-2} M^2
- 6.57\times 10^{-4} M^3
\end{equation}
($M_Z$ and~$M$ in solar masses).
We shall express our results for $Z\equiv M_Z/M$ in terms of the solar
metallicity $Z_\odot=0.02$.

\subsection{Stellar population synthesis}
\label{s:SPS}

Each star particle ($\sim 10^6\,M_\odot$) created during a run can be
regarded as a simple stellar population (SSP), i.e. a set of coeval
stars of uniform metallicity and known mass function.
We synthesize maps and spectra for a simulated galaxy by summing
the spectral energy distributions (SEDs) tabulated for SSPs by Bruzual
\& Charlot (1993), and so trace the magnitude and colour evolution of
the object.

\section{Runs and results}
\label{s:runs}

\subsection{Runs without star formation}
\label{s:nsf}

For both our halos, we first perform a simulation with only dark
matter particles.
We compare these results to those of runs with cooling gas but no
star formation (with baryon fractions $\Omega_b=0.05$ and~0.025; the
latter is expected to cool about twice as slowly by virtue of the
lower gas densities).

\begin{figure}[t]
\plottwo{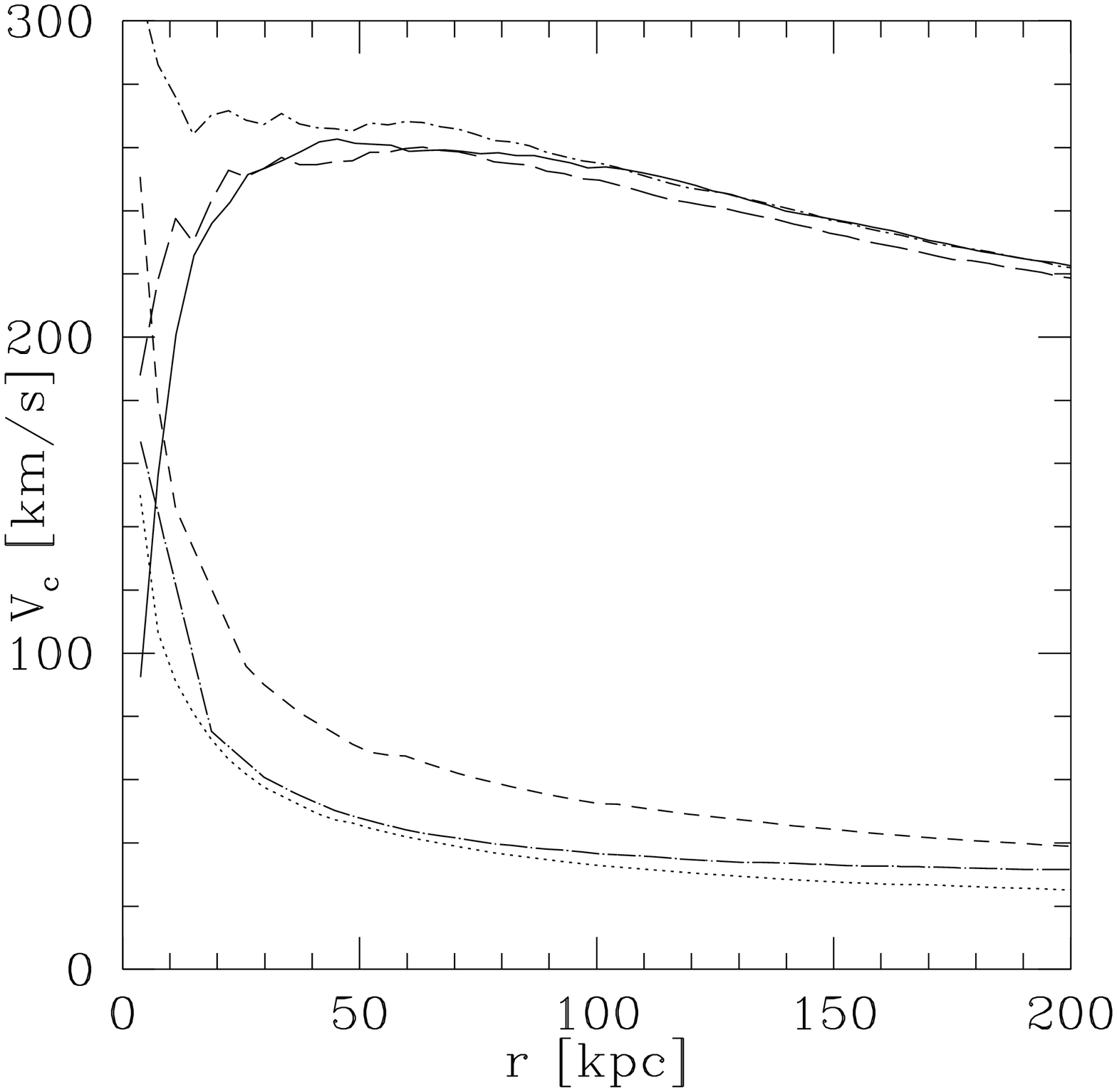}{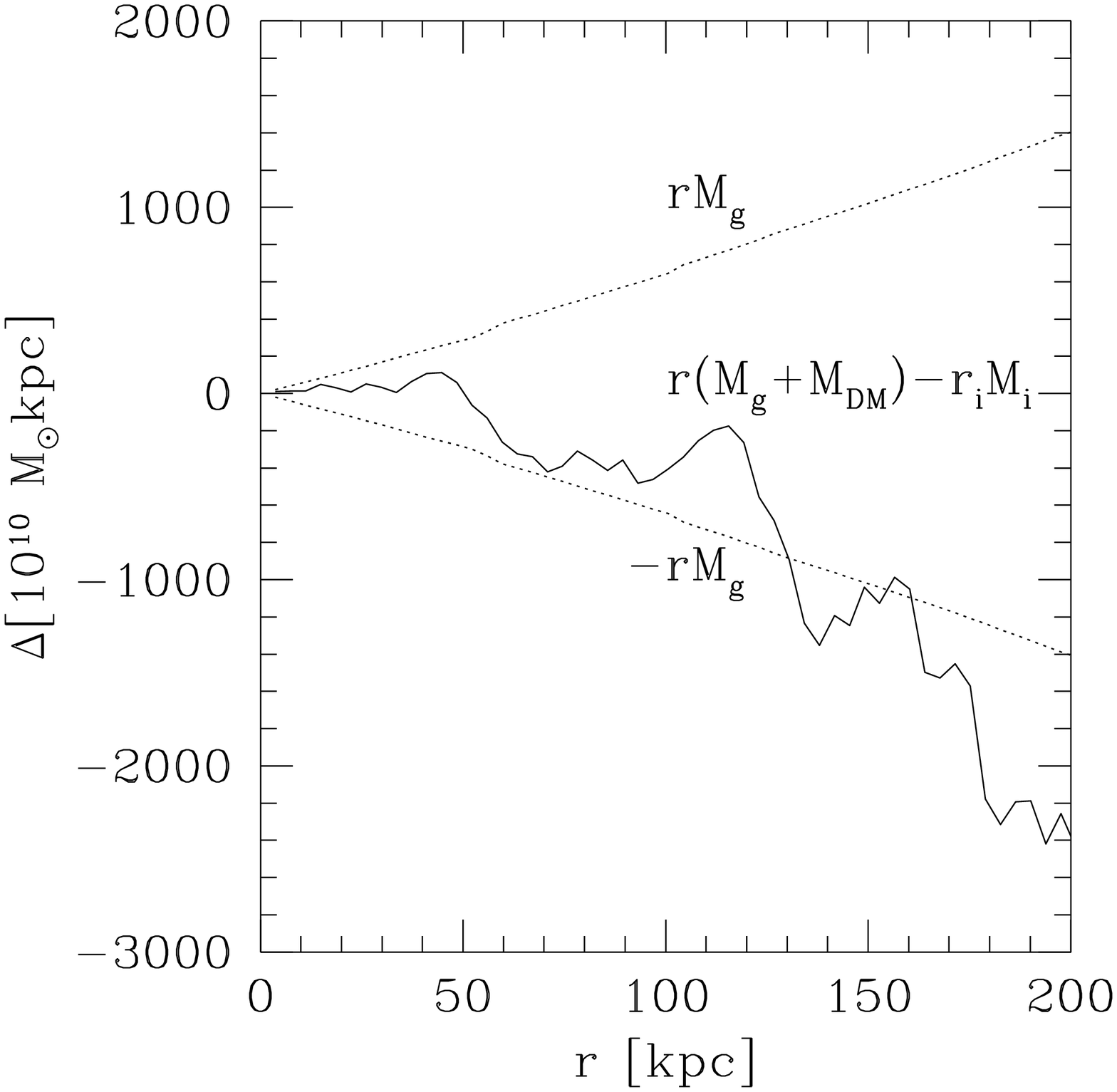}
\caption{\emph{Left:} rotation curve $V_c(r)\equiv GM(<r)/r$ at $z=0$
for runs with dark matter only (full curve), $\Omega_b=0.05$ (dark
matter, long dashes; gas, short dashes; total, dot-short dash),
$\Omega_b=0.025$ (gas, dotted curve), and $\Omega_b=0.05$ with UV
heating (gas dot-long dash curve).
\emph{Right:} demonstration of the adiabatic invariance of $r M(<r)$ by
comparing the radial profiles wirh dark matter only ($r_i M_i$) and
with $\Omega_b=0.05$ ($r(M_{\rm DM}+ M_{\rm gas})$). 
At $r\la 50$--100~kpc, the difference $r(M_{\rm DM}+M_{\rm
gas})-r_iM_i$ (solid curve) is 
small by comparison with $rM_{\rm gas}$ (dotted curves).
\label{f:b6nsf}}
\end{figure}

At $\Omega_b=0.05$ most of the gas within the virial radius condenses
at the centre, forming a disk of radius comparable to the
gravitational softening and causing the total rotation curve to rise
more than would be consistent with observations.
This cooling catastrophe is reduced by halving $\Omega_b$, as
expected.
Introducing a UV heating term has a similar effect at large radii, but
in addition suppresses the outer part of the central disk.
Confirming previous works, these results suggest that some form of
heating due to star formation is needed to form systems with a
distribution of baryons similar to that of observed spiral galaxies.
The left panel of figure~\ref{f:b6nsf}%
compares rotation curves for the gas component from these various runs.

The collapsing gas pulls along some of the dark matter, causing the
halo to become more centrally concentrated.
Previous studies (e.g., NFW) have relied on the adiabatic invariance
of $rM(<r)$ (Gunn 1977) to estimate this effect.
We tested this assumption by comparing the profiles for
our DM-only and cooling gas runs, and found that it holds:
at small radii ($r\la50$~kpc), where the adiabaticity
condition that the dynamical time be much shorter than the gas infall
time is satisfied, the difference between the $rM$ profiles is much
smaller than the smallest individual term ($rM_{\rm gas}$) in the
comparison.
This is illustrated by the right-hand panel of figure~\ref{f:b6nsf}.

\subsection{Runs with star formation}

\begin{figure}[t]
\plottwo{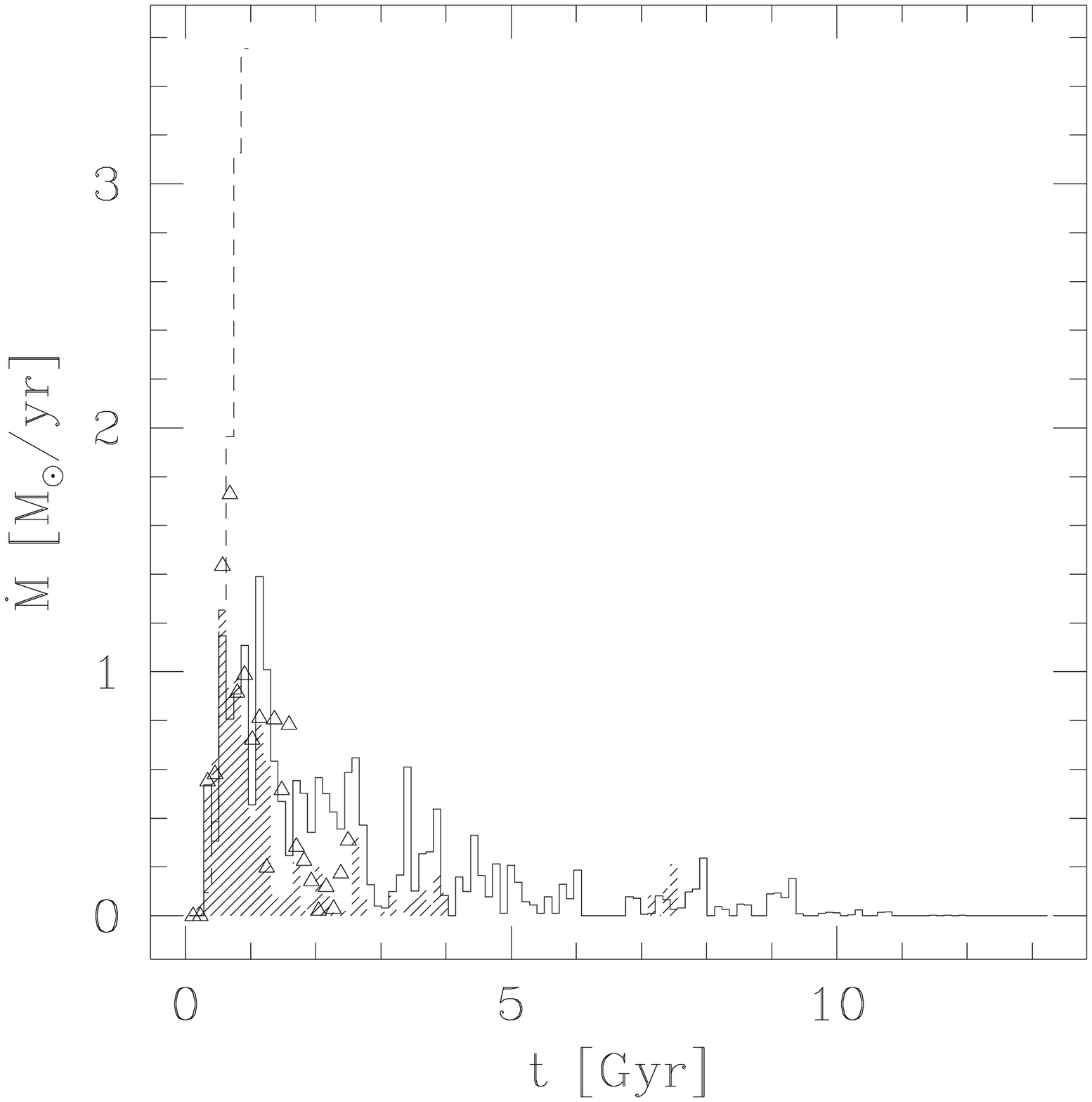}{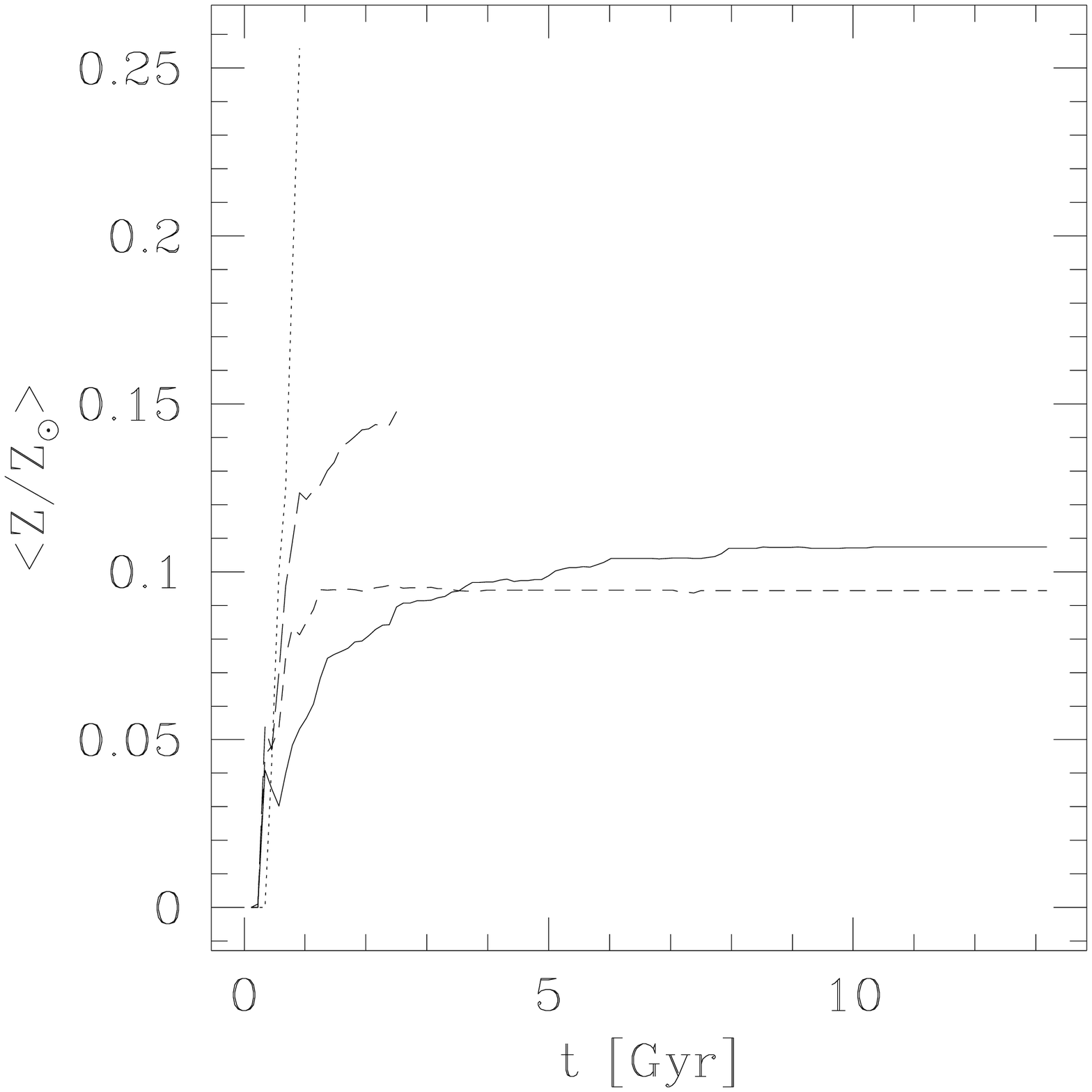}
\caption{
\emph{Left:}\label{f:sfr}
Star formation rate as a function of time since the start of
the simulation ($z=39$), for $V_c=120\,{\rm km}\,{\rm s}^{-1}$. 
Solid curve: no heating. All others include UV
heating. Shaded histogram: same star formation recipe. Triangles:
higher density threshold. Dashed curve: $t_s>t_d$ requirement
included.
\emph{Right:}\label{f:zc}
Cumulative mean metallicity of the stars in that same halo.
Solid curve: standard star formation recipe, no UV. Dashes: same
recipe, with UV. Long dashes: higher density threshold. Dotted curve:
$t_s>t_d$.
}
\end{figure}

For each halo we performed two sets of runs
using our fiducial star formation recipe, one with and the other
without UV heating.
We also tested alternative recipes on the small halo.
Figure~\ref{f:sfr}(left) shows the star formation rates for the
smaller halo.
The ones for the other halo are an order of magnitude larger (10
vs.~1~$M_\odot\,{\rm yr}^{-1}$),
and exhibit the same decay with a time scale of~$\sim 3$~Gyr
after the initial burst.
The rates are in the observed range for high-$z$ galaxies 
(Steidel et al 1996).
Star formation almost ceases when $z\la 0.5$.
The UV background further
suppresses star formation at late times, but is not
nearly as effective in the early phases when the density is still
high.

Raising the density threshold causes only a modest
increase in the star formation rate, showing that our results are not
critically sensitive to the exact value used.
A higher density threshold means that the supernova feedback
will reheat the gas to a lower temperature, allowing it to cool again
more quickly; this leads to greater recycling of the gas, which will
also be visible in the metallicity evolution.
Imposing the Jeans instability criterion (dashed curve) 
results in a spectacular increase in the star formation rate.
Our feedback is clearly too weak to prevent runaway burning of all the
available gas in this case.

Metallicity turns out to be a powerful discriminant between star formation
models.
Figure~\ref{f:zc}(right)
shows the mean metal content of all stars formed up to a
given time; the curves become horizontal when star formation stops.
(Results for the larger halo are $Z/Z_\odot\sim 0.3$--$0.4$.)
The UV background causes the metallicity to rise faster at first,
since the star formation is restricted to higher densities where the
feedback is less effective at mixing the material; but overall the
longer duration of the star formation phase without UV heating leads
to slightly higher metal fractions.
Raising the density threshold produces a moderate increase in the
metallicity, while the Jeans instability
criterion has more spectacular consequences, leading to uncomfortably
high mean $Z$ values for a small galaxy at high redshift.

By $z=0$ our objects have ``puffed up'' and acquired a very extended
envelope where old, metal-poor stars from accreted lumps dominate the
gas out to 150--200~kpc.
These might correspond to the dwarf spheroidal satellites of galaxies
like the Milky Way, but the resolution of our simulations does not
allow a firm identification.

\begin{figure}[t]
\plottwo{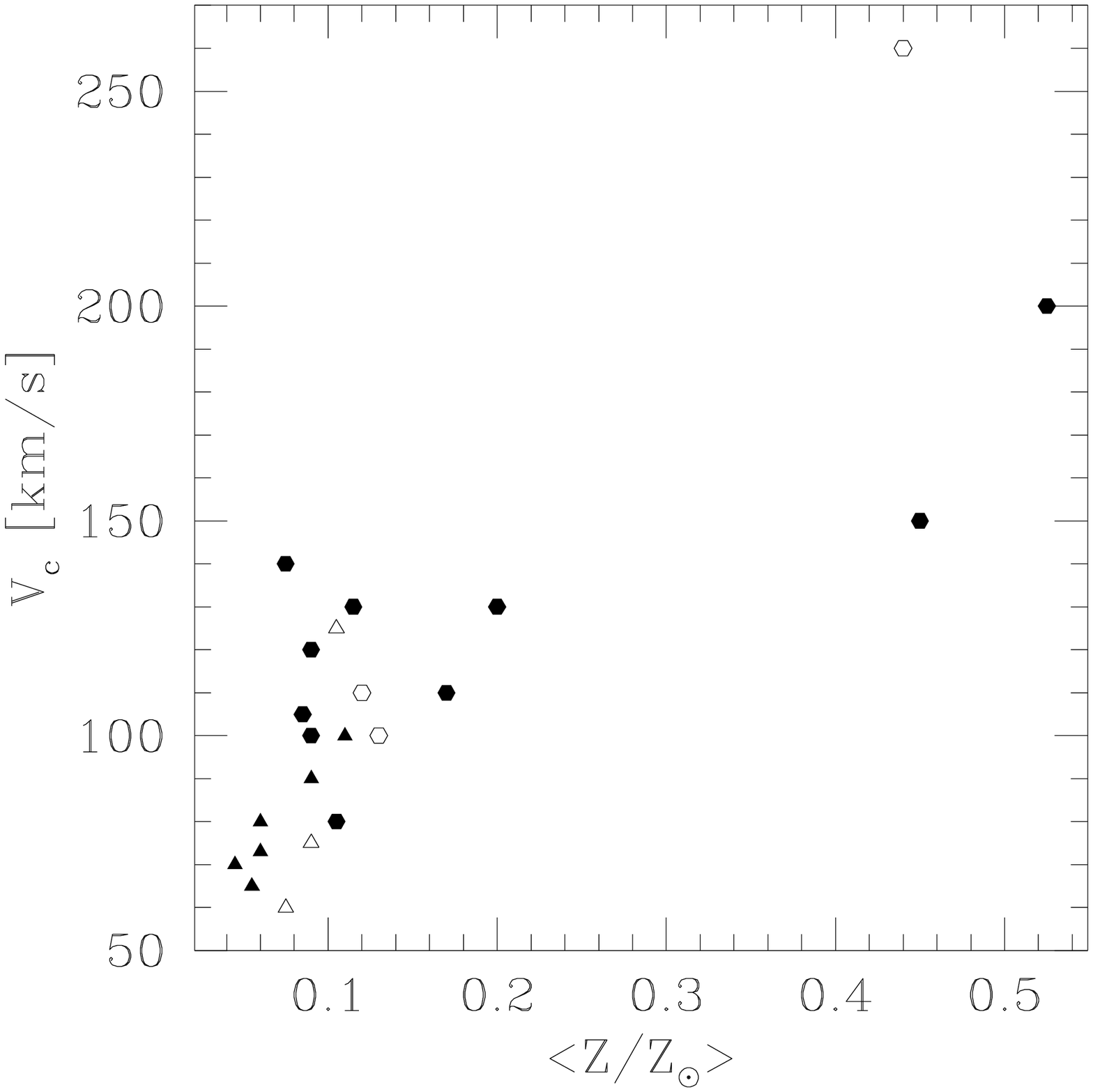}{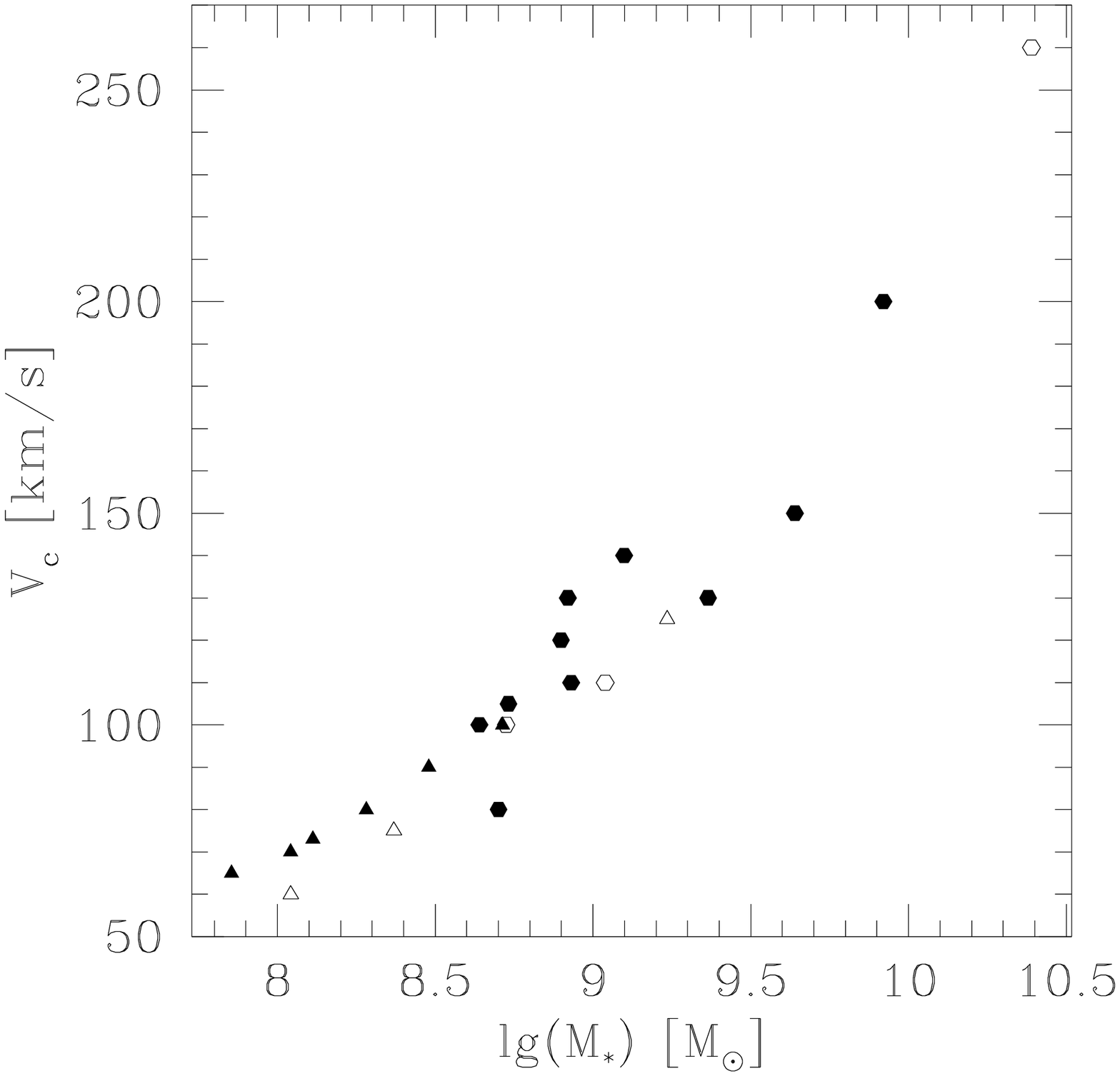}
\caption{
\label{f:vccorr}
Correlation between circular velocity $V_c$ and mean metallicity
(left) and mass in stars (right) for various sub-halos at $z\sim 2$
(filled symbols) and $z\sim 0.5$ (open symbols). 
Hexagons are from our larger B6 simulation, triangles from the smaller,
higher resolution B9 run.}
\end{figure}

It can be useful to study the individual lumps before they merge.
Figure~\ref{f:vccorr}
shows a tight Tully-Fisher-like correlation between the mass in stars
and the circular velocity, and a weaker but still clear link between
circular velocity and mean metal enrichment.
The correlations show little evolution between $z\sim2$ and $z\sim
0.5$, although naturally the mean $V_c$ does increase.
That data from both our simulations, with different mass resolutions,
fit essentially the same curve gives us some confidence that these
particular trends are not too sensitive to the details of our method.
Unfortunately, while the trend is qualitatively right, the slope of
the relation between the mass in stars and the circular velocity is
shallower than in the fits of Persic et al. (1996).
Thus we have not yet solved the well-known problem of forming too many
stars early on in small halos. This will probably require a better
understanding of what triggers star formation in primordial gas.
It has also been suggested (White \& Frenk 1991) that a stronger
supernova feedback may help cure this problem; we plan to investigate
this idea in more detail.

\begin{figure}[t]
\plottwo{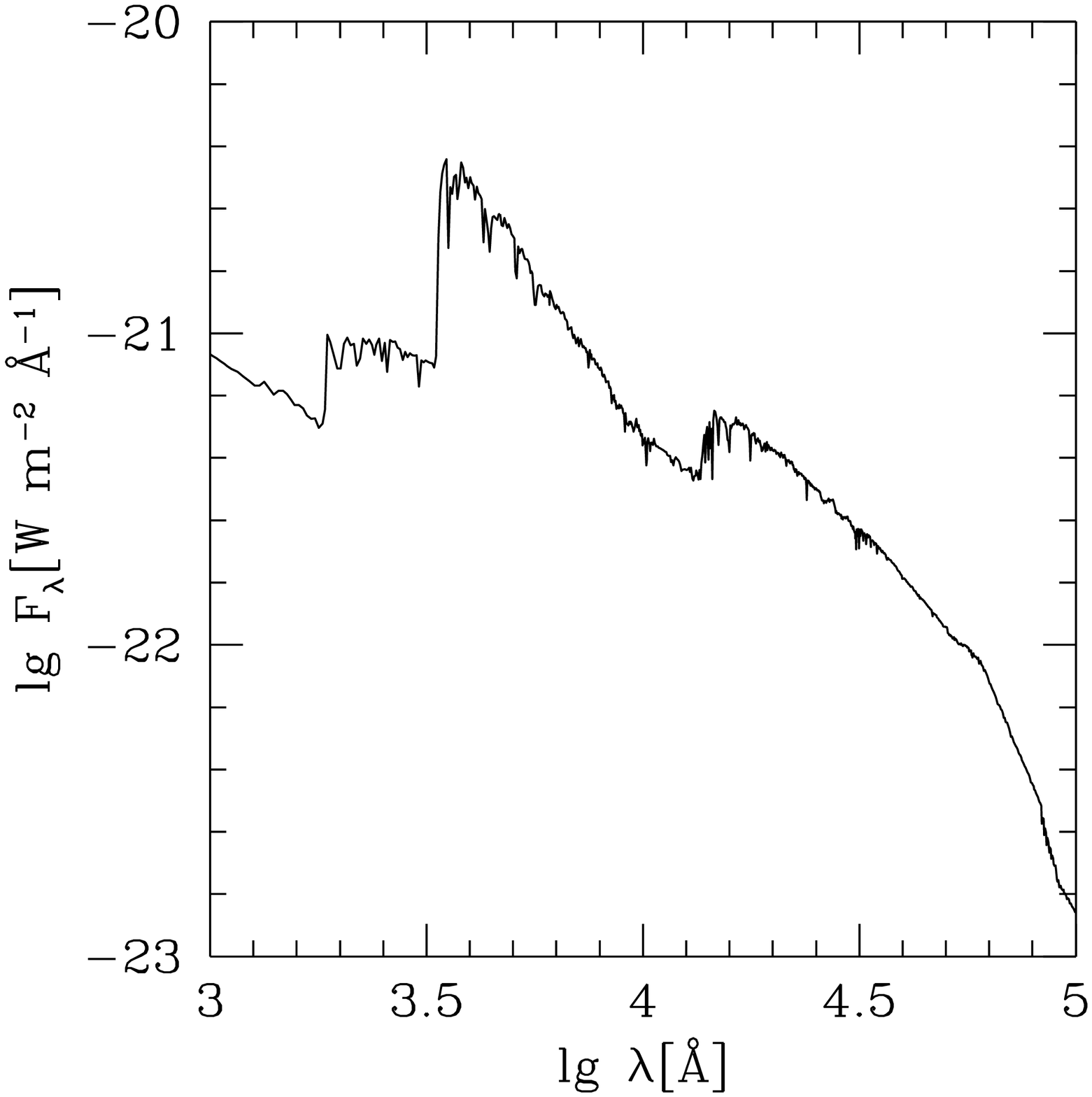}{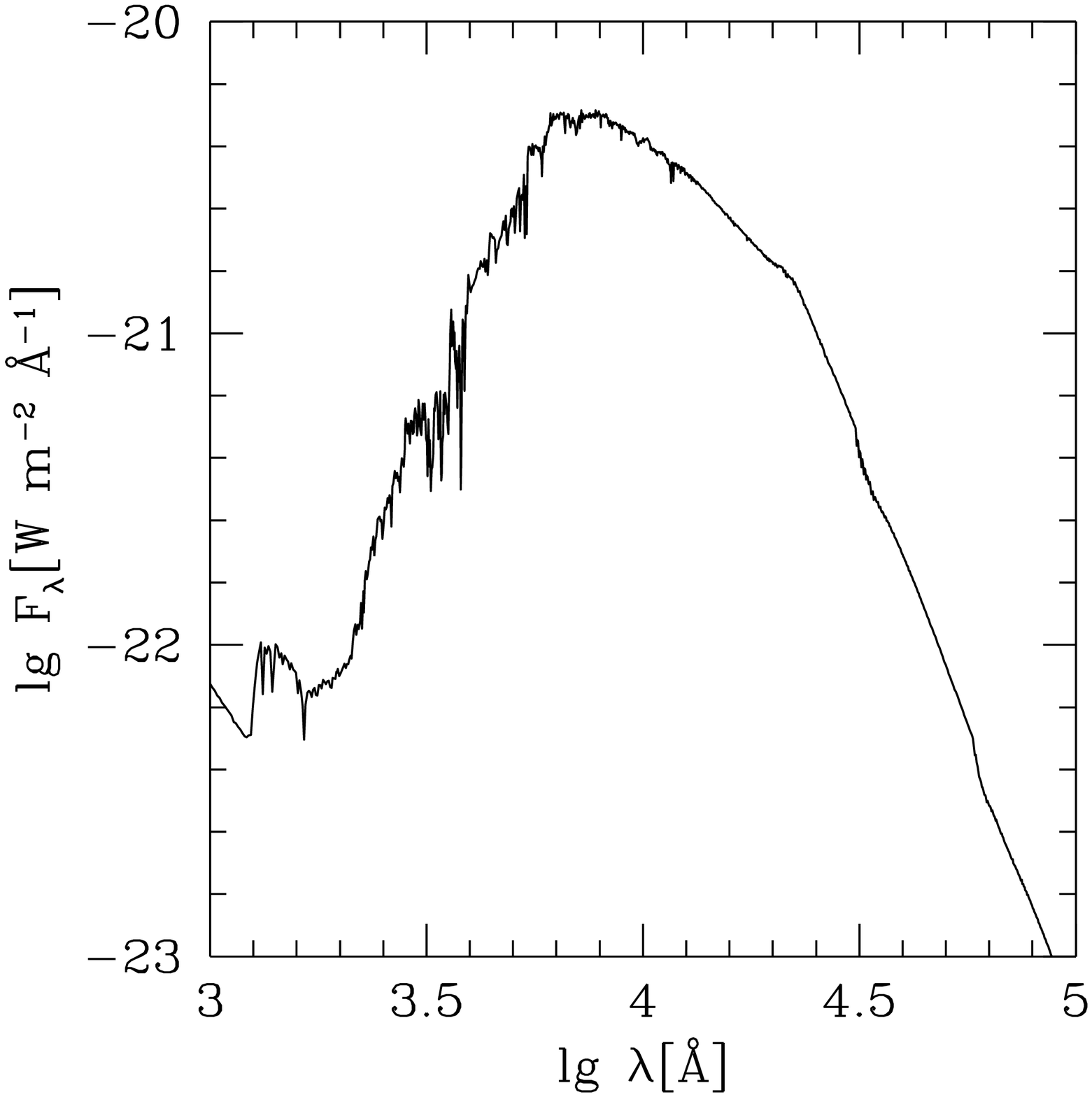}
\caption{
\label{f:sed}
Spectral energy distribution for the $V_c=120$~km/s halo at $z=2.64$
(left) and at $z=0.36$ (right). The abscissa is wavelength in the
observer's frame, the ordinate the flux per wavelength interval.
We assumed $\Omega=1$, $H_0=50$~km/s/Mpc.
}
\end{figure}

Finally, we present in figure~\ref{f:sed} the SED for our B9 halo
(fiducial recipe, with UV) at $z=2.64$ (left) and $z=0.36$
(right), synthesized with Bruzual \& Charlot's (1996) library which
takes account of the ages and metallicities of the individual SSPs.
The results have been translated into wavelengths and fluxes in the
observer's frame, but we haven't yet accounted for IGM absorption
(Madau 1995).
These two redshifts correspond respectively to a starburst phase and
to a time at which star formation has long since ceased.

\section{Discussion}

Some significant aspects of our simulation results are in good
agreement with observations.
The star formation peaks in a reasonable redshift range $2\la z \la 3$.
A related success is that the SEDs of the simulated galaxies at these
redshifts look much like those of observed objects.
Expected trends such as the relation between halo mass and star
formation rate, or between halo mass and metallicity, are reproduced
at least qualitatively, and we hope that the remarkable tightness of
the correlations will lead to a strong calibration of the star
formation prescription.
Such calibration is a necessary task, since plausible variants of the
prescription yield very different outcomes.
This is both a curse (it decreases our confidence in the more
quantitative aspects of the results) and a blessing, or at least a
promise that something useful may be learned about the star formation
and feedback processes.

The results at low redshift are less satisfactory.
The fact that we cannot sustain star formation to $z=0$ in objects
that we'd like to identify with spiral galaxies is a disappointment.
It could signal a flaw in the star formation algorithm.
It could also be due in part to our limited resolution, which favours
the formation of a hot, diffuse gas phase and may underestimate the
amount of cool gas available at late times to form a rotationally
supported disk of the appropriate size and mass.
Convergence studies will clearly be important in settling this issue.
In particular the dependence of the star formation and feedback
recipes on the numerical resolution deserves close attention.

The current description of the feedback is probably inadequate:
without incorporating some form of kinetic energy injection, the
heating is too weak to prevent runaway star formation when the gas
density is high, which will happen more easily in higher resolution
models.
As noted by NW (and confirmed by our own tests) the results are very
sensitive to the (arbitrary) magnitude of this kinetic energy
injection.
It would be useful to lift this arbitrariness with the help of studies
of the propagation of supernova winds in the ISM.

A difficulty should be noted in our implementation of the metallicity
evolution: we did not account for dilution of the heavy element
abundances by mixing in the ISM.
Consequently, we may be overestimating the scatter in stellar
metallicities, and in some models (with a high $\rho_{\rm min}$ and
weak feedback) the mean metallicity as well.

It appears from a scan of recent literature that the art of modelling
the cooling of gas within dark matter halos is approaching maturity,
with convergence properties and spurious heating effects being addressed.
Including star formation is the next step, and much progress is still
needed before we can have full confidence in our models' predictions.
Our preliminary results, uncertain as they may be, show some
encouraging similarities with observational data. They also exhibit
some rather tight relationships that should be eminently testable;
this last fact is reason enough to continue perfecting this simulation
technique.

\acknowledgments

We thank Julio Navarro for kindly making his SPH code available to us,
and Adriano Fontana, 
George Lake, Julio Navarro, Massimo Persic, Tom Quinn, Paolo
Salucci for stimulating discussions.


\begin{references}

\reference Bruzual, G., \& Charlot, S. 1993, \apj, 405, 538
\reference Giallongo, E., Cristiani, S., D'Odorico, S., Fontana, A.,
\& Savaglio, S. 1996, \apj, 466, 46
\reference Gunn, J.\ E. 1977, \apj, 218, 592
\reference Katz, N. 1992, \apj, 391, 502
\reference Katz, N., Weinberg, D.\ H., \& Hernquist, L. 1996, \apjs, 105, 19
\reference Kennicut, R. 1989, \apj, 344, 685
\reference Madau, P. 1995, \apj, 441, 18
\reference Monaghan, J.\ J. 1992, \araa, 30, 543
\reference Navarro, J.\ F., Frenk, C.\ S., \& White, S.\ D.\ M. 1996,
\apj, 462, 563 (NFW)
\reference Navarro, J.\ F., \& Steinmetz, M. 1996, preprint astro-ph/9605043
\reference Navarro, J.\ F., \& White, S.\ D.\ M. 1993, \mnras, 265, 271
(NW)
\reference Pardi, M.\ C., Ferrini, F., \& Matteucci, F. 1995, \apj,
444, 207
\reference Persic, M., Salucci, P., \& Stel, F. 1996, \mnras, 281, 27
\reference Steidel, C.\ C., Giavalisco, M., Pettini, M., Dickinson,
M., \& Adelberger, K.\ L. 1996, \apj, 462, L17
\reference Steinmetz, M., \& M\"uller, E. 1995, \mnras, 276, 549
\reference Weinberg, D.\ H., Hernquist, L., \& Katz, N. 1996, preprint
astro-ph/9604175
\reference White, S.\ D.\ M., \& Frenk, C.\ S. 1991, \apj, 379, 52
\reference Woosley, S.\ E., \& Weaver, T.\ A. 1995, \apjs, 101, 181
\end{references}
\end{document}